\documentclass[aps,pra,twocolumn,notitlepage,superscriptaddress]{revtex4-1}

\usepackage{graphicx,color}
\usepackage{dcolumn}
\usepackage{bm}
\usepackage{amsmath,amssymb,mathrsfs}
\usepackage{url}
\usepackage{hyperref}

\newcommand{\R}{\mathbb{R}}
\newcommand{\C}{\mathbb{C}}

\newcommand{\set}[1]{\mathsf{#1}}
\newcommand{\grp}[1]{\mathsf{#1}}

\def\d{{\rm d}}

\def\>{\rangle}
\def\<{\langle}
\def\kk{\>\!\>}
\def\bb{\<\!\<}
\newcommand{\st}[1]{\mathbf{#1}}
\newcommand{\bs}[1]{\boldsymbol{#1}}

\newcommand{\map}[1]{\mathcal{#1}}
\newcommand{\Tr}{\operatorname{Tr}}

\newcommand{\arccot}{\mathrm{arccot}\,}

\newcommand\myuparrow{\mathord{\uparrow}}
\newcommand\mydownarrow{\mathord{\downarrow}}

\begin{document}
	\preprint{APS/123-QED}
    \title{Benchmark for  the  quantum-enhanced control  of  reversible dynamics
    }
    \author{Yin Mo} 
    \affiliation{Department of Computer Science, The University of Hong Kong, Pokfulam Road, Hong Kong}
     \author{Giulio Chiribella} \affiliation{Department of Computer Science, The University of Hong Kong, Pokfulam Road, Hong Kong}
    \affiliation{Canadian Institute for Advanced Research, CIFAR Program in Quantum Information Science, Toronto, ON M5G 1Z8}
    \nopagebreak

	\begin{abstract} 
	    Controlling  quantum systems is crucial for  quantum computation and a variety of new  quantum technologies.  The control is typically achieved by breaking down the target dynamics into a sequence of elementary gates,  whose description can be stored into the memory of a classical computer.      Here we explore a different approach, initiated  by Nielsen and Chuang \cite{nielsen1997programmable},  where  the target  dynamics is encoded in the state of a quantum system, regarded as a ``quantum program".    	      We   show that quantum strategies based on coherent interactions between the quantum program and the target system  offer an advantage over all classical strategies that measure the program and conditionally operate on the system.    To certify the advantage,   we provide a benchmark that guarantees  the successful demonstration of quantum-enhanced programming   in realistic experiments. 
    \end{abstract}
    \maketitle

      \section{Introduction}   
      
      The ability to control quantum systems is at the core of quantum computing and  of a   new generation of quantum technologies \cite{dowling2003quantum}.  Most often, the control  is achieved 
      by decomposing the target dynamics into a sequence of elementary operations, whose execution can be controlled by a classical program, like a piece of code stored in the memory of a classical computer. This is the case, for example, in the circuit model of quantum computing and in digital quantum simulations.    A different approach was put forward by Nielsen and Chuang \cite{nielsen1997programmable}, who proposed that the dynamics of a quantum system could be  encoded in the state of an another quantum system. Such a state serves as a {\em quantum program}, containing the instructions needed to execute the target dynamics.  What makes the program quantum is that, in general, the instructions corresponding to two distinct dynamics can be encoded into two non-orthogonal quantum states.  Nielsen and Chuang's  paradigm led to the design of  programmable quantum gates \cite{hillery2002probabilistic,vidal2002storing,brazier2005probabilistic,hillery2006approximate,ishizaka2008asymptotic} and measurements \cite{fiuravsek2002universal,d2005efficient}, with  applications to quantum state discrimination \cite{duvsek2002quantum,bergou2005universal,sentis2013programmable}, quantum communication  \cite{bartlett2009quantum}, and quantum learning  \cite{bisio2010optimal, marvian2016universal}. 
Experimental demonstrations of programmable quantum devices have been reported  in a variety of setups  \cite{soubusta2004experimental,gopinath2005programmable,mivcuda2008experimental,bartuuvskova2008programmable,slodivcka2009experimental}, with applications to the experimental study of commutation relations \cite{yao2010experimental} and to the activation of entanglement \cite{adesso2014experimental}.    
        The quantum mechanical nature of the program  introduces  genuinely new questions.   How does the size of the program affect the accuracy? 
        How many times can one reuse the program before it loses its ability to specify  the target dynamics?  Is the best performance attained in a classical way, by reading out the program   and conditional operating  on the  data, or is it attained in a quantum way, by letting program and data interact as a closed system? 
        
\begin{figure}[ht]
        	\centering
        	\includegraphics[width=0.4\textwidth]{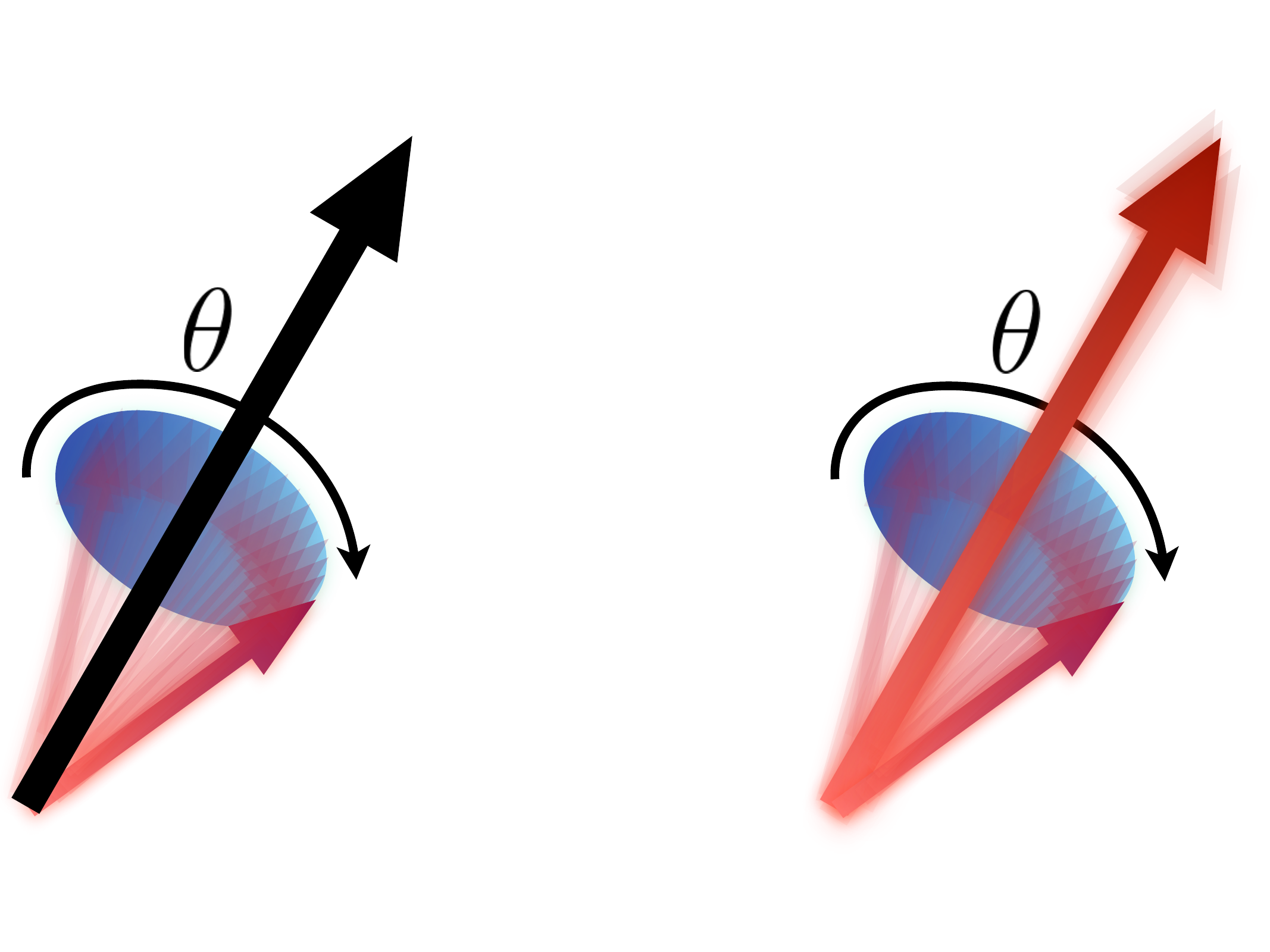}
        	\caption{\footnotesize
        		\textbf{Programming  rotations  with quantum spins.}  {\em On the left:}  the spin of a quantum particle (red arrow) is rotated by an angle $\theta$ around a well-defined classical axis (black arrow). {\em On the right:} The information about the rotation axis is encoded into the spin of a quantum particle, which serves as a quantum program controlling the target dynamics.}    
        	\label{fig1}
        \end{figure}

        In this paper   we answer these three questions, focussing on the problem of rotating the spin of a quantum particle around a direction determined by the  spin of another particle, as  illustrated in Figure \ref{fig1}. 
        We establish the ultimate quantum limit to the accuracy as a function of the size of the control spin, showing that error vanishes inverse linearly with the spin size.  The limit is  attained by a coherent  mechanism, whereby the  two spins  interact with one another  without leaking any information into the outside world.   We show that this coherent strategy     reaches a higher accuracy than  every  classical strategy where the control spin is measured in order to extract information about the program. 
    For example,  measurement-based strategies using a control system of spin $j=3/2$  can achieve at most fidelity  $64\%$ in  flipping a target qubit around a variable direction,  while the optimal quantum strategy achieves  $71\%$ fidelity.  
      The  gap between quantum and classical strategies allows us to establish a benchmark for the demonstration of quantum-enhanced programming  in realistic experiments: even if the implementation is affected by noise and imperfection, there is still a range of values of the fidelity within which the experiment can demonstrate a performance that could not be achieved classically, even with unlimited technology.  
        While many  benchmarks have been identified for quantum teleportation and cloning    \cite{hammerer2005quantum,adesso2008quantum,owari2008squeezing,namiki2008fidelity,calsamiglia2009phase,chiribella2013optimal,chiribella2014quantum,yang2014certifying},  to the best of our knowledge no benchmark for the task of programming quantum gates has been established prior to the present work.  
          
     The paper is structured as follows.   We start in Section \ref{sec:accuracysize} with the determination of the optimal tradeoff between accuracy and size of the quantum program in the case of qubit gates.  The corresponding quantum benchmark is then presented in Section \ref{sec:benchmark}.  
      We show how to surpass the benchmark, by       constructing a physical realization of the optimal programmable gate (Section \ref{sec:realization}) and showing that the advantage persists even if the program state is recycled multiple times (Section \ref{sec:longevity}).   In Section \ref{sec:larger}, we extend our result from qubits to systems of arbitrary dimension, proving that the advantage of quantum programming  is generic and providing an explicit analysis for the task of programming rotation gates.  The conclusions are finally drawn in Section \ref{sec:conclusions}.

\medskip 

 \section{Accuracy vs  size}\label{sec:accuracysize} 
 
     Suppose that a magnetic field is turned on for a limited period of time.  During this time, the direction of the field can be encoded  into the state of a magnetic material, which orients itself  along the  direction  of the field and ideally maintains  the orientation even after the field is turned off.   Thanks to this property, the magnet  can be used as a program  to  reproduce the dynamics that would have occurred {\em if a  particle were immersed in that field}.   This kind of reproduction can be seen as an elementary learning process, where a quantum machine  learns to emulate an unknown dynamics by observing its action on a certain input \cite{bisio2010optimal, marvian2016universal}. 
   Now, the question is: how large must the magnet be  in order to accurately steer the desired dynamics?  
 
The answer depends on the size of the particle one wishes to control.    Let us  focus first on the case of a single qubit, embodied in a spin-$1/2$ particle. 
   In this case, the target dynamics is a rotation around the direction of the field, of an angle proportional to the time of the evolution. The rotation is represented by the matrix 
        $ V_{\theta, \bf n}    =     \exp [-i  \theta  \,   {\bf   n}  \cdot {\bs \sigma}/2]\, ,$
       where   $\theta$ is the rotation angle, depending on the strength of the field and on the evolution time,   $ {\bf n }  =  (n_x,n_y,n_z)$ is the rotation axis, corresponding to the direction of the field,  and   ${\bf   n}  \cdot {\bs \sigma}  =  n_x \sigma_x +  n_y\sigma_y +  n_z\sigma_z$  is a linear combination of Pauli matrices, representing the projection of the spin operator along the direction $\st n$. The small magnet  is modelled as a spin-$j$ particle, whose state $|\phi_{\st n}\>$ serves as an indicator of the direction $\st n$, and as a program for the rotation gate $V_{\theta,\bf n}$.   We impose  that the encoding $\st n\mapsto |\phi_{\st n}\>$ is consistent with the physical interpretation of $\st n$ as a  spatial direction. This means that rotating the direction $\st n$  should be equivalent to  rotating the state vector---in formula,
\begin{align}\label{phin}
|\phi_{g  \, \st n}  \>   =   U_g^{(j)}  \,  |\phi_{\st n}  \>  \, ,    
\end{align} 
where $g$ is an arbitrary rotation in three dimensional space and $U_{g }^{(j)}$ is the  unitary matrix representing the action of  $g$ on the Hilbert space of the spin-$j$ particle. 
   Except for Eq. (\ref{phin}),  we make no assumption on the program states.

        Once the information about the rotation axis is  encoded in the control spin,  the problem is to devise a  mechanism that emulates  rotations  around that axis.    Mathematically, the mechanism is described by a quantum channel \cite{nielsen2000quantum}, describing the joint evolution of the control and target.  
         To evaluate the accuracy of the control mechanism, we compare the output state of the target  with the ideal  output of the   gate $V_{\theta, \bf n}$.   As a figure of merit, we use the fidelity
            \begin{align}\label{F}
        	    F(j,\theta,  {\bf n},  \psi)=\<\psi|V_{\theta,\bf n}^{\dag}  \,  \left[  \mathcal{C}_{\theta} \left (\phi_{\bf n}\otimes\psi  \right)\right ]  \,  V_{\theta,\bf n}|\psi\> \ ,
            \end{align}
           where $|\psi\>$ is the initial state of the data qubit and $\map C_\theta$ is the quantum channel describing the effective evolution from the  control and target together  to the target alone. 
     Note that {\em a priori}  the fidelity could depend on the input state $|\psi\>$ and 
     on the rotation axis $\st n$. 
  To eliminate the dependence, one can consider 
 the {\em average input-output fidelity}  \cite{gilchrist2005distance}
           \begin{align}\label{Fave}
        	    F(j,\theta) =  \int d {\bf n}  \,  \int d\psi 
        	    \,  F(j,\theta,\bf n ,  \psi) \, , 
            \end{align}
        where   $d \bf n$ is the invariant probability distribution on the unit sphere and  $d \psi$ is the  invariant probability distribution on the pure states of the system.    In actual experiments, the averages over all directions and over all states can be replaced by averages over a finite set of directions and states, using the theory of unitary designs \cite{dankert2009exact}.      

\begin{figure}[ht]
        	\centering
        	\includegraphics[width=0.48\textwidth]{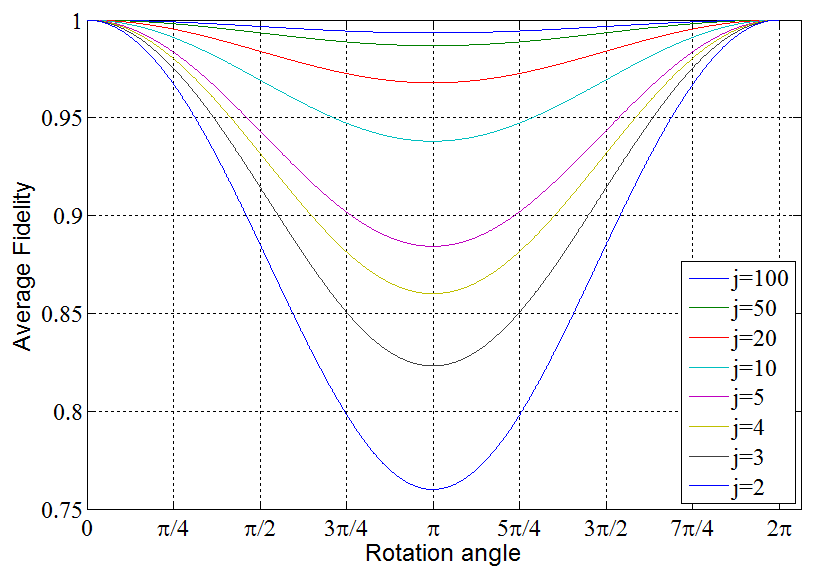}
        	\caption{\footnotesize
        		\textbf{Average fidelity for  different rotation  angles.}   The dependence of the fidelity on the rotation angle $\theta$ is illustrated for different values of the spin from $j=2$ to $j=100$.   The fidelity is minimum for $\theta  =  \pi$, meaning that rotations of $180$ degrees are the hardest to program. }
        	\label{fig2}  
        \end{figure}

  Our first result is the optimal  quantum scaling of the  fidelity with the program size. 
    By maximizing over the quantum channel $\map C_\theta$ and over the program states $|\phi_{\st n}\>$ we find the  optimal value
    
            \begin{align}\label{Fopt}
        	    F_{\rm opt} (j,\theta)\nonumber=&\dfrac{1}{3}+\dfrac{2}{3(1+2j)^{2}}\bigg[2j^{2}+\dfrac{2j+1}{2}+\dfrac{2j+1}{2}\cos\theta  \\
        	    & ~~ +j\sqrt{1+2(2j+1)\cos\theta+(2j+1)^{2}}\bigg]  \,, 
            \end{align}
            valid for $j \ge 3/2$  (see Appendix \ref{app:A} for the derivation and  for the expression of the fidelity in the $j=1/2$ and $j=1$ cases).
        The dependence  on the rotation angle is illustrated in Figure \ref{fig2}, where one can see that the fidelity  is minimum for  $\theta  =  \pi$, meaning that   the rotations of $180$ degrees are the hardest to program.    
    
        Eq.(\ref{Fopt}) gives the exact expression of the fidelity, but an even more insightful expression can be obtained by Taylor expansion, which yields the approximate formula 
            \begin{align}\label{Foptapp}
    	        F_{\rm opt}(j,\theta)  =  1-\dfrac{1-\cos\theta}{3j}+
    	        O\left(\dfrac{1}{j^2}\right) \, .
            \end{align}
This result shows that the error (defined as 1 minus fidelity)  tends to zero as the control spin becomes macroscopically large. Note that the scaling  $1/j$ refers to the average over all possible rotation axes and over all possible input states.  Later, we will prove that the scaling $1/j$ is optimal even in the worst-case over all input states.

\section{Benchmark for coherent quantum control}\label{sec:benchmark} We have established the ultimate quantum performance in programming qubit rotations. An important question is whether this performance can be achieved through  a classical strategy where the  program  is measured  and a conditional operation  on the data is performed.  We refer to these strategies as \emph{measure-and-operate (MO) strategies}. 
  In Appendix \ref{app:optMO}  we determine  
the maximum fidelity achievable by arbitrary  MO strategies, providing a benchmark that can be used to certify the demonstration of quantum-enhanced programming in realistic experiments.  
    \begin{figure}[ht]
    	    \centering
    	    \includegraphics[width=0.47\textwidth]{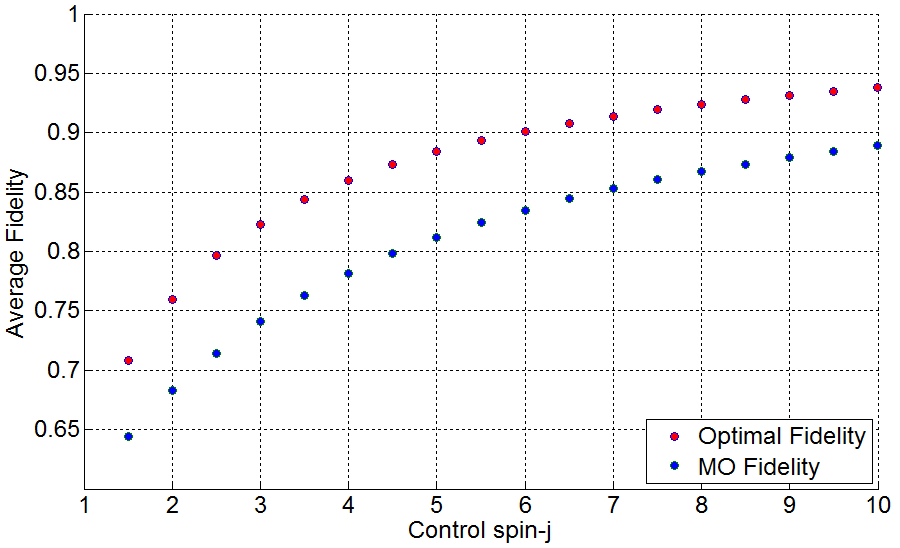}
    	    \caption{\footnotesize
    		    \textbf{Benchmark for coherent quantum control.}   The quantum benchmark (blue dots) and the optimal quantum fidelity (red dots) are plotted for rotations of $180$ degrees in a function of the spin size, with $j$ ranging from $3/2$ to $10$. }
    	    \label{fig:benchmark}
        \end{figure}
     The exact value of the benchmark, derived in Appendix \ref{app:optMO}, is 
         \begin{align}
         \nonumber F_{\rm  MO}&(j,\theta)    =\dfrac{4j+4+(2j+1)\cos(\theta-\tau)}{6j+9} \\
         &~~~~~~~~~ +\dfrac{(2j+1)(\cos\theta+\cos\tau)+\cos(\theta+\tau)+1}{3(j+1)(2j+3)} \, ,
     \end{align}
     with $\tau  =  \arccot    [(2j^{2}+3j+2)\cos\theta+2j+1]/[(2j^{2}+3j)\sin\theta]$.
       Figure \ref{fig:benchmark}  shows the gap between the benchmark and the optimal quantum fidelity  for small values of the control spin. 
    For large spins, the benchmark takes the asymptotic value
             \begin{align}\label{FMOapp}
        	    F_{\rm MO} (j,\theta) = 1-\dfrac{2(1-\cos\theta)}{3j} 
        	    +O\left(\dfrac{1}{j^2}\right) \, .
            \end{align}
Note that  the error (one minus fidelity)   is exactly twice the error of the  optimal quantum protocol, which can be read out from  Eq.(\ref{Foptapp}).  The error goes to zero both for quantum and classical strategies, but the rate for quantum strategies is twice as fast.   Intriguingly, this seems to be a recurring feature in the relation between optimal classical and quantum programming strategies, as we will see later in the paper.  
 

\medskip 
\section{Physical realization of the optimal programmable gate}\label{sec:realization}
        We have seen that the optimal quantum strategy offers an advantage over all  classical, measurement-based strategies.   
        But how  is the advantage achieved, concretely?  
     For the program states, we find out that the optimal choice is to use  spin-coherent states \cite{arecchi1972atomic}, encoding the direction $\st n$ in the spin coherent state with maximum projection along the direction $\st n$.   This is not unexpected, because spin coherent states are  optimal for the  estimation of spatial directions  \cite{holevo2011probabilistic}. Still,  estimation and programming are two distinct operational tasks---and indeed, estimation is {\em not} the optimal strategy for controlling  rotations.  It is a non-trivial  open question whether    any deeper connection exists that links  the optimal setup for  quantum  estimation with the optimal setup for quantum control.  
       
Regarding the control mechanism,  we find that  the channel $\map C_\theta$ has an \emph{economical} realization \cite{fuchs1997optimal, niu1999two, durt2005economical}, meaning that it  can be implemented by letting the program and the data  interact as a closed system,  without introducing  extra ancillas. 
        Explicitly, the optimal quantum channel is realized through the unitary evolution 
            \begin{align}\label{optimalU}
        	    U_\theta  =     \exp  \left[    -i f(\theta) \, \frac{  \,   \st J \cdot \bs \sigma}{2j+1}  \right]  
            \end{align}
        where $\st J  =  (J_x, J_y,  J_z)$ are the spin operators of the control spin, $\st J \cdot \bs \sigma   = \sum_{i=x,y,z} J_i\otimes  \sigma_i$ is the Heisenberg coupling, and   $ f(\theta)$ is the function 
            \begin{align}\label{ftheta}
        	     f(\theta)  &=      \arccos \frac{  1  +  (2j+1) \cos \theta }{\sqrt{ [1+  (2j+1)  \cos \theta ]^2 + [(2j+1)  \sin \theta]^2 }}  \, .
            \end{align}
            
            Physically, this unitary evolution can be realized by setting up an isotropic spin-spin  interaction, described by the Hamiltonian $H  =   \alpha  \, \st J \cdot \bs \sigma$, for some suitable coupling constant $\alpha$, and by letting the two spins evolve for  time 
            \begin{align}t (\theta)  =       \frac{f(\theta)}{(2j+1)  \alpha \hbar}\, ,
            \end{align}
             depending on the  angle  $\theta$ of the target rotation. Remarkable,  the same program states and the same interaction can be used to control the full time evolution of the target system:  one has only to adjust  the interaction time [determined by  the angle $f(\theta)$] based on the evolution time in the target dynamics [determined by the angle $\theta$]. 
        
        Eq.(\ref{optimalU}) shows that the optimal way to program the dynamics  is to set up  a Heisenberg interaction between the control and target spins. This result answers an open question raised by Marvian and Mann  \cite{marvian2008building}, who assumed the Heisenberg interaction  and showed that it can be used to approximate arbitrary rotations in the limit of large $j$ limit.  Marvian and Mann asked whether the Heisenberg interaction achieves the best scaling of the error with the spin size, a question that is answered in the affirmative by our result.   The optimality of the Heisenberg interaction is not limited to the average fidelity: in terms of scaling with $j$,  the unitary gate (\ref{optimalU}) is optimal also in the worst case over all input states.  Indeed, one can explicitly evaluate the worst case fidelity, which takes the value   
 \begin{align}\label{Fwc}
        	    F_{\rm w} (j,\theta) = 1-\dfrac{1-\cos\theta}{j} 
        	    +O\left(\dfrac{1}{j^2}\right) \, .
            \end{align}
Note that the error scaling $1/j$ is the best  one could have hoped for, because  the average  error is a lower bound to the worst case error and we know from Eq. (\ref{Foptapp}) that the  average error cannot vanish faster than   $1/j$. 

Knowing the value of the worst case fidelity, one can estimate the error  scenarios where one or more  gates are  implemented with a quantum program.  For example, a quantum circuit that uses  $k$ programmable single-qubit gates will have an error of size $k/\sqrt j$ with respect to the ideal functionality.  This means that, in order to have a negligible error, the size of the program should be large compared to the square of the number of programmable gates in the circuit. 

 \medskip 
 
\section{Longevity of the quantum advantage}\label{sec:longevity} 

The Heisenberg interaction transfers  information  from the program to the  data.     This leads to  a backreaction effect, whereby the program gradually  loses its ability to control operations on the target \cite{bartlett2006degradation}.  An important question is how many times the program  can be reused before the accuracy drops below a certain threshold.  The number of reusing times was called  \emph{longevity}  in Ref. \cite{bartlett2006degradation}.  Another important question is how many times the control spin can be reused before the  quantum advantage is lost.   
We will refer to this number as the \emph{longevity of the quantum advantage}.

Suppose that the joint evolution of  control and target is described by the same unitary gate  at every step. Assuming the gate to be of the form  of Eq.  (\ref{optimalU}) for  some fixed function $f(\theta)$,  we obtain the   close-form  expression 
             \begin{align}
            F(j,\theta,n) = 1 - \dfrac{1-\cos\theta}{3j}\cdot\dfrac{n(1-\cos\theta)+j}{j}
            \end{align}
            quantifying the average fidelity at the leading order in $j$ (see Appendix \ref{app:longevity} for the derivation). 
         From this expression one can see that the longevity grows as $j^2$.  
            However, the longevity of the quantum advantage is much shorter: comparing the above fidelity with the MO fidelity in Eq.(\ref{FMOapp}), we  find that the quantum advantage disappears if the number of repetitions is larger than 
             \begin{align}
            L   (j,\theta)=  \dfrac{j}{1-\cos\theta} + O\left(1\right)    \, .
            \end{align}
One can also consider  more elaborate strategies  where the interaction time between control and target  is optimized at every step.    However, these strategies do not increase the longevity of the quantum advantage in the large $j$ limit. 
    
    

\medskip 

        \section{Programming larger systems}\label{sec:larger}   Our result establishes the existence of a quantum advantage  for  single-qubit gates.  This finding  is conceptually important, because the advantage for single qubits implies an advantage  of coherent programming for quantum systems of arbitrary  dimension. Indeed, one can  immediately prove the advantage by using the qubit benchmark for gates that act nontrivially only in a fixed two-dimensional subspace.
        
Our results also  give a heuristic for the control of higher dimensional spins.   
        The idea is to encode the rotation axis in a  spin coherent state  and to let  the control and target spin interact as closed system. Explicitly, we make two spin systems undergo the Heisenberg interaction $U_\theta^{(k)}  =     \exp  \left[    -i \theta \,   \,   2\st J \cdot \st K/ (2j+1)  \right]$,   
         where $\st K  =  (K_x, K_y,  K_z)$ are the spin operators of the target spin.  
        Using the unitary gate $U_\theta$, in Appendix \ref{app:spink} we obtain the average fidelity 
              \begin{align}
                	F(j,k,\theta)  =   1-\dfrac{k(2k+1) (1-\cos\theta)}{3j}    \, ,         
             \end{align}
        in the large $j$ limit.  Remarkably, the error grows \emph{quadratically}---rather than linearly---with the size of the target spin: in order to ensure high fidelity, the size of the program  must be large compared to the square of the size of data.  The same conclusion holds for the worst case fidelity, which has the asymptotic  expression  
        \begin{align}
        F_{\rm w}(j,k,\theta)  =  1  - \frac  { [k(k+1)  +  c (k)]\,  (1-\cos\theta)}{j}    \,, 
        \end{align} 
       with $c(k)  = 0 $ for even $k$ and $c(k)=  1/4$ for odd $k$.

The quantum strategy exhibits an advantage over the MO strategy consisting in measuring the direction $\st n$ from the spin coherent state pointing in direction $\st n$  and performing a rotation based on the outcome. 
    Again, we find that the error of the quantum strategy vanishes in the macroscopic limit of large control systems, at a rate twice as fast  than  the error of the classical strategy.    

 \medskip

\section{Conclusions}\label{sec:conclusions}    

   We determined the ultimate accuracy for   the execution of rotations controlled by quantum spins.   The ultimate accuracy limit is achieved through a Heisenberg  interaction, with the interaction time depending on the rotation angle and on the spin size.   Our work calls for the experimental realization of programmable setups that achieve the ultimate quantum limits to the control of  rotation gates.   For small values of the spin, a possible testbed is provided by NMR systems, where spin-spin interactions are naturally available \cite{vandersypen2005nmr}.   
        Another possibility is to use quantum dots, where one can engineer a coupling between a  single spin and an assembly of spins effectively behaving as a single spin $j$ particle  \cite{chesi2015theory}. This scenario, named the \emph{box model}, can be achieved through a uniform coupling of a central spin to the neighbouring sites.  
        No matter what platform is used,  our results provide the  rigorous benchmark that can be used to validate the successful demonstration of quantum-enhanced programmable gates.

On the fundamental side, our work unveils a deep relation  between the classical and quantum approaches to programming.    In the classical approach, the target gate is approximated by a sequence of elementary gates, whose number grows as $\log 1/\epsilon$ with the error parameter $\epsilon$  \cite{kitaev1997quantum}. At the leading order, the number of gates is equal to the number of bits used by the classical program that  describes the target gate.  In the quantum approach, we found that the target gate is approximated with error $\epsilon  =  O(1/j)$, implying  that the number of program qubits needed to achieve error $\epsilon$ scales  as $  \log(2j+1)   =O(\log 1/\epsilon)$.   
        In other words, our result shows that the classical and quantum approaches are asymptotically equivalent  in terms of tradeoff between accuracy and size. It is worth noticing, however, that our quantum strategies were constructed from a symmetry assumption, namely that the rotations in space are reflected into rotation of the program states [cf. Eq. (\ref{phin})]. Although physically reasonable, this assumption may be lifted, by allowing arbitrary encodings of the target dynamics in the program states.  Removing the symmetry assumption  (\ref{phin}) might  in principle lead to an improved size-accuracy tradeoff beyond the $O(1/j)$ scaling observed in this paper.    Determining whether this is the case  requires the development of new techniques beyond the  scope of the present investigation. While we expect the scaling $O(1/j)$ to be remain optimal even with general encodings, we believe that the development of  techniques to tackle the question  has the potential to   reveal new relations between quantum programming, quantum metrology, and quantum simulations. 
        
 \medskip 
  
{\bf Acknowledgements.}         This work is supported by the Hong Kong Research Grant Council through Grant No. 17326616 and 17300317, by National Science
        Foundation of China through Grant 11675136, by the HKU Seed Funding for Basic Research, and by the Canadian Institute for Advanced Research (CIFAR). 

    \bibliographystyle{apsrev4-1}
    \bibliography{ProgrammingRotations}

\appendix

 \section{Optimal quantum fidelity}\label{app:A}
        The fidelity $F(j,\theta)$ in Eq.(3) of the main text is the result over two averages: the average over all pure states and the average over all rotations axes. Quite conveniently, the average over the states can be eliminated by using the well-known relation with   the \emph{entanglement fidelity} \cite{horodecki1999general}.  With the notation of our paper, the relation reads
        \begin{align}\label{horodecki}   
        	\int  d\psi  \,  F(j,\theta,  {\bf n} ,  \psi)  =\dfrac{1}{3}+\dfrac{2}{3} \, F^{\rm (e)}(j,\theta,{\bf n})  \, ,
        \end{align}
        where  $ F^{\rm (e)}(j,\theta,{\bf n})$ is the entanglement fidelity, given by
        \begin{align}
        	F^{\rm (e)}(j,\theta,{\bf n})    =     \<\Phi^+_{\theta, {\bf n}}|  \,\left[   \left(  \map C_\theta  \otimes \map I\right)  \left(   \phi_{\bf n} \otimes   \Phi^+   \right)\right]  \,   |\Phi^+_{\theta,\bf n}\> \, .
        \end{align}
        Here $\Phi^+$ denotes the projector on the canonical maximally entangled state  $|\Phi^+\>   =  (|0\>|0\>   +  |1\>|1\>)/\sqrt 2 $ 
        and  $|\Phi^+_{\theta,{\bf n}}  \>$    is the rotated maximally entangled state defined by 
        $ |\Phi^+_{\theta,{\bf n}}  \>:   = (  V_{\theta, \bf n}  \otimes I)   |\Phi^+\>  \, . $
        
        Using Eq.(\ref{horodecki}), the average fidelity can be rewritten as
            \begin{align}\label{F-Fe}
            	F(j,\theta)=\dfrac{1}{3}+\dfrac{2}{3} F^{\rm (e)}(j,\theta) \, ,
            \end{align}
        where $F^{\rm (e)}(j,\theta)$ is the average entanglement fidelity
            \begin{align}\label{Feave}
            	F^{\rm (e)}(j,\theta) =\,  \int d {\bf n} \,\<\Phi^{+}_{\theta,{\bf n}}|  \, (\mathcal{C}_{\theta}\otimes \mathcal{I})(\phi_{\bf n}\otimes\Phi^{+}) \, |\Phi^{+}_{\theta,{\bf n}}\> \ .
            \end{align}
        
      The fidelity can be conveniently rewritten using the ``double ket notation" 
        \begin{align}
            |\Psi\kk:=\sum_{m}\sum_{n} \<m|\Psi|n\>|m\>|n\> \, .
        \end{align} 
        Denoting by  $C_{\theta}$ the Choi operator for channel $\mathcal{C_{\theta}}$, we obtain
            \begin{align}\label{Feave-change}
            	F^{\rm (e)}(j,\theta)&=\nonumber\dfrac{1}{2}\int\d {\bf n}\Tr\left[C_{\theta}\,|\overline{\phi}_{\bf n}\>\<\overline{\phi}_{\bf n}|\otimes|\Phi^+_{\theta,{\bf n}}\>\<\Phi^+_{\theta,{\bf n}}|\right]\\
            	&=\dfrac{1}{4}\<\overline{\phi}|\bb V_{\theta}|C_{\theta}^*|\overline{\phi}\>|V_{\theta}\kk \, ,
            \end{align}
having defined
            \begin{align}
                C_{\theta}^*:=\int\d {\bf n}\left({U_{\bf n}^{(j)T}}\otimes U_{\bf n}^{\dag}\otimes U_{\bf n}^{T}\right)C_{\theta}
                \left(\overline{U}_{\bf n}^{(j)}\otimes U_{\bf n}\otimes \overline{U}_{\bf n}\right) \, .
            \end{align}    
                    
        Furthermore, it is convenient to define the operator
         \begin{align}
        	\widetilde C_{\theta} = (\sigma_{y}^{(j)}\otimes I\otimes\sigma_{y}) \ C_{\theta}^* \ (\sigma_{y}^{(j)}\otimes I\otimes\sigma_{y}) \ . 
        \end{align}
With this definition, it is easy to prove the relation 
        \begin{align}
            \left[\widetilde C_{\theta}, U_{g}^{(j)}\otimes U_{g}\otimes U_{g}\right]=0,\qquad \forall g\in SU(2)\, .
        \end{align}   
         
         Now, using Schur's lemma  we obtain the expression
            \begin{align}\label{Choi}
            	\widetilde C_{\theta}=\alpha P_{j+1}\oplus\beta P_{j-1}\oplus P_{j}\otimes M_{j} \ ,
            \end{align}
        where $P_l$ is the projection on the subspace with total angular momentum $l$,  $\alpha, \beta, \gamma$ are complex coefficients, and  $M_{j}=\begin{pmatrix}
        \gamma_{\rm A} & \gamma_{\rm B} \\ \gamma_{\rm C} & \gamma_{\rm D}
        \end{pmatrix}$ is a non-negative matrix. 
        The trace-preserving condition on the channel is equivalent to the constraint 
             \begin{align}\label{Choi-coefficient}
            	\Tr_{\rm target}[C_{\theta}]=&I_{\rm control}\otimes I_{\rm target} \ ,
        \end{align}   
        on the Choi operator. In terms of the coefficient, this implies  the condition	
	    \begin{align}\label{Choi-coefficient'}
    \begin{cases}
            		\dfrac{2j+3}{2j+2}\alpha+\dfrac{2j+1}{2j+2}\gamma_{\rm A}=1\\  \\ 
            		\dfrac{2j-1}{2j}\beta+\dfrac{2j+1}{2j}\gamma_{\rm D}=1    \, .       	\end{cases}
            \end{align}
        
        Now, we insert the Eqs. (\ref{Choi}) and (\ref{Choi-coefficient'})  into the expression of the fidelity [Eq. (\ref{Feave-change})]. 
         After a long calculation using Clebsch-Gordan coefficients we can get the expression 
             \begin{align}\label{Feave-Shur}
            	F^{\rm (e)}(j,\theta)=\nonumber&A\sin^{2}\frac{\theta}{2}+B\cos^{2}\frac{\theta}{2}\\
            	&+C\overline{J}_{z}\sin\frac{\theta}{2}\cos\frac{\theta}{2}+D\overline{J_{z}^{2}}\sin^{2}\frac{\theta}{2} \ ,
            \end{align}
        where $\overline{J}_{z}$   ($\overline{J_{z}^{2}}$) is  the  average (of the square of the) $z$-component of the  angular momentum, while the constants $A, B, C, $ and $D$ are as follows:
             \begin{align*}
            	\begin{cases}
            		A=&\dfrac{1}{2(1+2j)}\Big[(j+1)\alpha+j\beta\Big]
            		\\B=&\dfrac{1}{2(1+2j)}\Big[(j+1)\gamma_{\rm A}+j\gamma_{\rm D}\\
            		&\qquad -\sqrt{j(j+1)}(\gamma_{\rm B}+\gamma_{\rm C})\Big]
            		\\C=&\dfrac{i}{2\sqrt{j(j+1)}}(\gamma_{\rm B}-\gamma_{\rm C})
            		\\D=&\dfrac{1}{2(1+2j)}\left[-\dfrac{\alpha}{j+1}-\dfrac{\beta}{j}\right.\\
            		&\qquad \left.+\dfrac{\gamma_{\rm A}}{j+1}+\dfrac{\gamma_{\rm D}}{j}+\dfrac{\gamma_{\rm B}+\gamma_{\rm C}}{\sqrt{j(j+1)}}\right]\, ,
            	\end{cases}
            \end{align*}
For $j\ge 3/2$, taking into account that the maximum expectation value  $\overline{J}_{z}$ is equal to $j$ and optimizing over the coefficients $\alpha$ and $\beta$ by $\gamma_{A}$ and $\gamma_{D}$ we obtain the optimal fidelity    
         \begin{align}\label{Feopt}
            	F_{\rm opt}^{\rm (e)}(j,\theta)=\nonumber&\dfrac{1}{(1+2j)^{2}}\left[2j^{2}+\dfrac{2j+1}{2}+\dfrac{2j+1}{2}\cos\theta\right.\\
            	&+\left.j\sqrt{1+2(2j+1)\cos\theta+(2j+1)^{2}}\right] \ . 
            \end{align}
    Note that the maximization of the expectation value $\overline{J}_{z}$ requires the program state to be the  spin-coherent state $|j,j\>$. 

   Eq. (\ref{Feopt}) gives the optimal value of the entanglement fidelity.    The optimal value of the average fidelity can then be obtained from Eq.(\ref{F-Fe}), which yields 
            \begin{align}\label{equ-A8}
            	F(j,\theta)_{\rm opt}=\nonumber&\dfrac{1}{3}+\dfrac{2}{3(1+2j)^{2}}\left[2j^{2}+\dfrac{2j+1}{2}+\dfrac{2j+1}{2}\cos\theta\right.\\
            	&+\left.j\sqrt{1+2(2j+1)\cos\theta+(2j+1)^{2}}\right] \ .
            \end{align}
        
\begin{figure}[ht]
    	\centering
    	\includegraphics[width=0.48\textwidth]{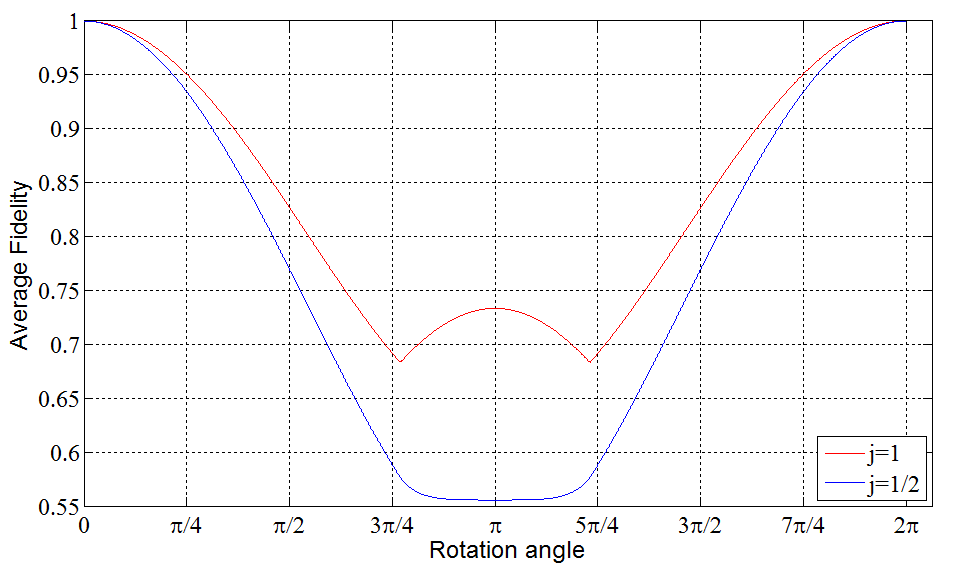}
    	\caption{\footnotesize
    		\textbf{Optimal quantum fidelity for $j=1/2$ and $j=1$.}  For $j=1/2$, a transition  occurs when $|\theta  -  \pi  |  = 2 \arctan \sqrt{4+\sqrt 7}$.   For $j=1$,  the transition occurs when $|\theta - \pi|    = 23/100\pi $. }
    	\label{fig:spin12}
    \end{figure}

The cases of $j=1/2$ and $j=1$ must be treated separately. In these two cases the optimal fidelity exhibits  critical points, as illustrated  in Figure \ref{fig:spin12}.

  For $j=1/2$, the optimal fidelity is
    \begin{align}
		    	F_{\rm opt}(j=\frac{1}{2},\theta)=	\dfrac{1}{3}+\dfrac{1}{6}\left[\dfrac{2+7\cos\theta}{6+12\cos\theta}-\dfrac{\cos\theta}{2}\right]	\end{align}
			for $|\theta - \pi  |\le    2  \arctan \sqrt{4+\sqrt 7}$, and 	\begin{align}
    	F_{\rm opt}(j=\frac{1}{2},\theta)=
    	\dfrac{1}{3}+\dfrac{3+2\cos\theta+\sqrt{5+4\cos\theta}}{12}
		\end{align}
		otherwise.   
    	For $j=1$, the transition happens for $|\theta -\pi| \le \delta$, with $\delta  \approx 0.23\,  \pi $. The optimal fidelity is 
		\begin{align}
	F_{\rm opt}(j=1,\theta) =  \dfrac{1}{3}+\dfrac{2\sin^{2}\theta}{5}
    	 		\end{align}
	for $|\theta-\pi|\le \delta $, and
	\begin{align}
    	F_{\rm opt}(j=1,\theta)=
    	\dfrac{1}{3}+\dfrac{7+3\cos\theta+\sqrt{10+6\cos\theta}}{27}  	\end{align}
  otherwise.  
 Quite surprisingly,  
  optimal program state for  $|\theta-  \pi|\le  \delta $ the is not the spin coherent state, but rather a $p$-orbital  $|1,0\>_{\st n}$ pointing in the direction of the  rotation axis.
	\section{Optimal measure-and-operate fidelity}\label{app:optMO}
   
    In a generic MO strategy, the measurement is described by a Positive Operator-Valued Measure (POVM) $\{P_x\}_{x\in\set X}$, where $\set X$ is the set of all possible outcomes  and $P_x$ is the positive operator associated to the outcome $x$. The conditional operation is described by quantum channel $\map C_{\theta,x}$ acting on the data qubit. 
    On average over all possible outcomes, the action of the classical protocol is represented by the MO channel 
        \begin{align}\label{CMO}
            	{ \map C}_{\theta, {\rm MO}}   (\rho)   =  \sum_{x\in \set X}  \, \Tr \left[P_x\,  \phi_{\bf n}  \right]  \,  \map C_{\theta, x}   ( \rho)   \, . 
        \end{align} 
    When the data qubit is in the state $|\psi\>$, the fidelity between the output of the MO channel and the desired output is 
            \begin{align}\label{FMO}
            	F_{\rm MO}(j,\theta,  {\bf n},  \psi)=\<\psi|V_{\theta,\bf n}^{\dag}  \,  \left[   {\mathcal{C}}_{\theta, {\rm MO}} \left ( \phi_{\bf n} \otimes \psi \right)\right ]  \,  V_{\theta,\bf n}|\psi\> \ .
            \end{align}
    Now, let $F_{\rm MO}  (j,\theta)$  be the average of the fidelity over all  input states and  over all possible rotation axes.    Using the relation   with the entanglement fidelity, we obtain    
        \begin{align}\label{equ-C3}
                	F_{\rm MO}(j,\theta)=\dfrac{1}{3}+\dfrac{2}{3}F^{\rm (e)}_{\rm MO}(j,\theta) \ ,
                \end{align}
        with
                \begin{align}
                	F^{\rm (e)}_{\rm MO}(j,\theta)=\sum_x \, \int  \d \st n   \frac{\bb   V_{\theta,\st n}  |  C_{\theta, x} |V_{\theta,\st n  }\kk \,    \<  \phi_{\st n}|   P_x |\phi_{\st n}\> }4     \, , 
                \end{align}
              where $C_{\theta, x}$ is the Choi operator of the channel $\map C_{\theta,x}$. 
  The entanglement fidelity can be rewritten as 
     \begin{align}
          \nonumber       	F^{\rm (e)}_{\rm MO}(j,\theta)  &=\sum_x \, \int  \d g   \frac{\bb   V_{\theta}    |\,  \left(  \map U_g \otimes \overline{\map U}_g \right)  (C_{\theta, x}) \,  |V_{\theta }\kk  }4 \,    \\  
          &  \qquad  \qquad\qquad    \times  \<  \phi | \,\map U_g^{(j)}  (P_x)\, |\phi \>     \, .  
                \end{align}
At this point, we define the  operators \begin{align}
P_g^{(x)}   :  =  (2j+1) \,\map U_g^{(j)}  (P_x)/\Tr[P_x] 
\end{align} and we note that they satisfy the normalization condition 
\begin{align}
\int \d g  \,   P_g^{(x)}   =  I    \qquad \forall x  \in \set X \, .
\end{align}  
In other words, the operators $\{  P_g^{(x)} \}$ define a POVM with outcome $g$. 
Similarly, we define the operators  $ C_g^{(x)}  =   \left(  \map U_g \otimes \overline{\map U}_g \right)  (C_{\theta, x}) $ and note that each of them is the Choi operator of a quantum channel  $\map C_g^{(x)}$.    
   Hence, the  POVM with operators $\{  P_g^{(x)} \}$ and the conditional channels $\map C_g^{(x)}$ form an MO strategy.  The entanglement fidelity can then be rewritten as 
    \begin{align}
 	F^{\rm (e)}_{\rm MO}(j,\theta)   =\sum_x  \,  p_x  \,          	F^{\rm (e)}_{\rm MO}(j,\theta,x)  \, ,
\end{align}
where $p_x$ is the probability $p_x:  =  \Tr[P_x]/(2j+1)$ and $	F^{\rm (e)}_{\rm MO}(j,\theta,x) $ is the fidelity of the $x$-th MO strategy.  
Hence, we  obtain the bound  
   \begin{align}
          \nonumber       	F^{\rm (e)}_{\rm MO}(j,\theta)   &\le \max_x  \,           	F^{\rm (e)}_{\rm MO}(j,\theta,x) \\
      \nonumber     &  =  (2j+1)\, \max_{C, \rho, \phi}        \int  \d g   \frac{\bb   V_{\theta}    |    C_g   |V_{\theta }\kk \,   \<  \phi | \rho_g |\phi \>  }4        \, ,
\end{align} 
where the maximization runs over all  Choi operators $C$ representing qubit channels and all density matrices $\rho$ representing spin-$j$ states,    and we defined    $C_g =  \left(  \map U_g \otimes \overline{\map U}_g \right)  (C_{\theta, x})$ and  $\rho_g  = \map U_g^{(j)}  (P_x)$. 

Using the relation  $\overline U_g  =    \sigma_y  \,  U_g \,  \sigma_y $, the bound on the fidelity becomes   
   \begin{align}\label{ultima}
   F^{\rm (e)}_{\rm MO}(j,\theta)   &\le  (2j+1)   \max_{C, \rho, \phi} \Tr [  (  C\otimes \rho) \,  \Omega]   \, ,
   \end{align} 
with 
\begin{align}\label{Omega}
     \Omega    =   \int  \d g    \left( \map U_g \otimes \map U_g  \otimes \map U_g^{(j)}  \right) \left(\frac{ |   W_{\theta}  \kk \bb  W_{\theta }|}4  \,   \otimes |\phi \>\<\phi|  \right)        
\end{align}
and   $W_\theta  =  \cos\frac \theta 2  \,    \sigma_y  +   \sin\frac \theta 2  \, \sigma_x$.  
Note that the maximization can be restricted to pure states, of the form $\rho  =  |\psi\>\<\psi|$.

    Now, the vector    $|W_\theta\kk$ can be rewritten as 
\begin{align}
   |W_\theta\kk  =  \sqrt 2  \,  \left[ i   \cos\frac \theta 2  \,  |0,0\>  + \sin\frac \theta 2  \,  |1,0\>   \right]  \, ,
\end{align}
having used the notation $  |l,m\>$ for the eigenstates of the $z$-component of the total spin.   Using this fact, we obtain that every vector $(  U_g\otimes U_g)  \,  |W_\theta\kk$ can be expanded in the basis $ \set B   =  \left\{    i  |0,0\>  ,   |1,0\>_x, |1,0\>_y,  |1,0\>_z\right\}
$, with 
  \begin{align} \nonumber     |1,0\>_z   & =     |1,0\>   \\   \nonumber |1,0\>_x  &=    \frac{ |1,1\>  +  i  |1,-1\>}{\sqrt 2}   \\ |1,0\>_y &=   \frac{|1,1\>   -  i  |1,-1\>  }{\sqrt 2}  \, ,
  \end{align} 
  and all the expansion coefficients are real. 
   Hence, the Choi operator  $C$ in Eq. (\ref{ultima}) can be chosen to have real matrix elements in the same basis.    Moreover, we can restrict the maximization to the Choi operators $C$  that are extreme points of the convex set of Choi operators with real matrix elements in this basis.      Using Choi's characterization of the extreme points  \cite{choi1975completely,bengtsson2007geometry}, we find out that the extreme real Choi operators are rank-one---that is, they represent unitary gates. 

Explicitly, we can write the Choi operator as  $C  =  |V_\tau\kk\bb V_\tau|$, with  
\begin{align}
\nonumber |V_\tau\kk   & = \sqrt{2} \left[ i  \cos \frac \tau 2  \,    |0,0\>  \right. \\
  & \left.  ~+   \sin \frac\tau 2   \,  \big(  r_x  \, |1,0\>_x +  r_y  \, |1,0\>_y +  r_z  \, |1,0\>_z   \big)\right] \, ,      
\end{align} 
where $\tau$ is an angle and $\st r   =  (r_x,r_y,r_z)^T$ is a unit vector in $\R^3$. Now, there must exist a rotation $h$ that transforms  the vector $\st r$ into the $z$ axis.  For this particular rotation, we have 
\begin{align}
\nonumber (  U_h \otimes U_h)  \, |V_\tau\kk  &  =   \sqrt{2} \left[   i  \cos \frac \tau 2  \,    |0,0\>   +    \sin \frac\tau 2 \,    |1,0\> \right] \\
&  =  |W_\tau\kk \, .     
\end{align} 
Since the operator $\Omega$ in Eq. (\ref{Omega}) is invariant under rotations, the bound on the fidelity becomes 
\begin{align}
F^{\rm (e)}_{\rm MO}(j,\theta)   &\le (2j+1)  \max_\tau \max_{|\psi\>}  ~ \bb   W_\tau |  \<\psi|  \Omega  |  W_\tau\kk  |\psi\>  \, .       
\end{align}

Now, note that we have 
\begin{align}
\nonumber \bb W_\tau  |     \,  U_g\otimes U_g  \,    |W_\theta\kk   &=     2\cos \frac \tau 2      \cos \frac \theta 2   \\
& \quad  +      2\sin \frac\tau 2      \sin \frac\theta 2   \,    \<1,0|      U^{(1)}_g  |1,0\>  \, .
\end{align}
and 
\begin{align}
\nonumber &\Big | \bb W_\tau  |     \,  U_g\otimes U_g  \,    |W_\theta\kk\Big|^2   \\
\nonumber &=     4\cos^2 \frac \tau 2      \cos^2 \frac \theta 2   +  \frac 43  \sin^2 \frac\tau 2      \sin^2  \frac\theta 2   \\
\nonumber & \quad  + \frac 83 \, \sin^2 \frac\tau 2      \sin^2  \frac\theta 2      \,     \< 2,0|   U_g^{(2)}  | 2,0\>  \\
   &  \quad +  8   \cos \frac \tau 2      \cos \frac \theta 2      \sin \frac\tau 2      \sin \frac\theta 2   \,    \<1,0|      U^{(1)}_g  |1,0\>  \, .
\end{align}
Using the above relation, we obtain 
\begin{align}
\nonumber &\bb   W_\tau |  \<\psi|  \Omega  |  W_\tau\kk  |\psi\>   \\
\nonumber &  =  \int \d g \,    \Big | \bb W_\tau  |     \,  U_g\otimes U_g  \,    |W_\theta\kk\Big|^2   \,  \Big|\<\psi  |  U_g^{(j)}|\phi\>\Big|^2\\
  &  =      \<\psi  |   \<\widetilde \psi |    \Gamma  |  \phi\> |\widetilde \phi\>  \, ,    
\end{align}
with  $|\widetilde \psi\>  :  =    e^{-i \pi    J_y } \,  |\overline \psi\>  $ and
\begin{align}\label{MOCG}
\nonumber \Gamma    &=   \Big( 4\cos^2 \frac \tau 2      \cos^2 \frac \theta 2   +  \frac 43  \sin^2 \frac\tau 2      \sin^2  \frac\theta 2\Big) \,  \Pi_{00}  \\
 \nonumber   & \quad  \frac 8{15} \, \sin^2 \frac\tau 2      \sin^2  \frac\theta 2  \,  \Pi_{20}  \\
   &  \quad    +   \frac 83   \cos \frac \tau 2      \cos \frac \theta 2      \sin \frac\tau 2      \sin \frac\theta 2 \,  \Pi_{10} \, , 
\end{align}
where we used the notation $\Pi_{jm}  =  |j,m\>\<j,m|$. 
Note that the projectors $\Pi_{j0}$ are invariant under multiplication with rotations around the $z$-axis, namely 
\begin{align}
\Pi_{j0}  =  \Pi_{j0}   \,(U_h\otimes U_h )  =   (U_h\otimes U_h )  \, \Pi_{j0}  \, , 
\end{align}
where $h$ is an arbitrary rotation $h$ around the $z$-axis. Hence, we have the bound 
  \begin{align}
\nonumber    \<\psi  |   \<\widetilde \psi |    \Gamma  |  \phi\> |\widetilde \phi\>   & \le \max_{m,m'} ~ (-1)^{m-m'}  \\
&  \quad \times    \<j,m  |   \< j,-m|   ~\Gamma  ~|  j,m'\>   |  j,  -m'\> 
  \end{align}  
 By direct inspection, we find that the above expression reaches its maximum for $m=m'=j$. Moreover, we find that the maximum over the angle $\tau$ is attained for
       \begin{align}\label{tauinapp}
      \cot\tau = \dfrac{(2j^{2}+3j+2)\cos\theta+2j+1}{(2j^{2}+3j)\sin\theta}, \qquad \theta \in [0,\pi] \, .
      \end{align}
      
For this value of $\tau$, we obtain the maximum fidelity 
      \begin{align}
      \nonumber F_{\rm  MO}^{(\rm e)}&(j,\theta)   = \dfrac{(2j+1)(1+\cos(\theta-\tau))}{2(2j+3)} \\
      & +\dfrac{(2j+1)(\cos\theta+\cos\tau)+\cos(\theta+\tau)+1}{2(j+1)(2j+3)}  \, .
      \end{align}
In terms of  average input-output fidelity, we obtain  the value
      \begin{align}
      \nonumber   &  F_{\rm  MO}(j,\theta)_{\rm opt}    = \dfrac{4j+4+(2j+1)\cos(\theta-\tau)}{6j+9} \\
  \label{FMO}    & +\dfrac{(2j+1)(\cos\theta+\cos\tau)+\cos(\theta+\tau)+1}{3(j+1)(2j+3)}  \, .
      \end{align}
      The maximum fidelity is achieved by using the program state $U_{g(\textbf{n})}|j,j\>$, measuring   the coherent state POVM $\{P_{{\textbf n'}}=U_{g({\textbf n'})}|j,j\>\<j,j|U_{g({\bf n'})}^{\dagger}\}$, and rotating around the axis $\bf n'$ of the angle $\tau$ determined by  Eq. (\ref{tauinapp}).   Note that the angle $\tau$ converges to $\theta$ in the large $j$ limit.

	\section{Longevity of the quantum advantage}\label{app:longevity}
        The state of the control spin after the interaction can be obtained by application of the complementary channel $ \widetilde{\map C}_\theta $, defined by 
            \begin{align}\label{recycle}
                \widetilde{\map C}_\theta   (\rho^{(j)})   =    \Tr_{\rm target} \left[     U_\theta   \,   \left ( \rho^{(j)} \otimes  \frac I 2  \right)   U^\dag_{\theta} \right]  \, .   
            \end{align}
        To evaluate this state, it is convenient to look at the evolution of the basis states $ |j,m\>_{\st n}$.  By explicit calculation, we obtain the relation  
            \begin{align}
                \widetilde{\map C}_\theta  \Big( |j,m\>\<j,m|_{\st n}   \Big)   =  \sum_{i=-1}^{1}  c_{m+i,m}    \,   |j,m+i\>\<j,m+i |_{\st n}   \ ,
            \end{align}
            where the coefficients $c_{m+i,m}$ are given by 
            \begin{align}
            	\begin{cases}
            		c_{m-1,m}=&\nonumber\dfrac{(j+m)(1+j-m)}{(1+2j)^{2}}(1-\cos\theta-\frac{\sin^{2}\theta}{2j})\\
            		c_{m,m}=&\nonumber 1-c_{m-1,m}-c_{m+1,m}\\
            		c_{m+1,m}=&\nonumber\dfrac{(j-m)(1+j+m)}{(1+2j)^{2}}(1-\cos\theta-\frac{\sin^{2}\theta}{2j})
            	\end{cases} \ ,
            \end{align}
    
        At the first step, the program starts in the state $|j,j\>_{\st n}$.  By repeatedly applying Eq.(\ref{recycle}), we then obtain the program state at every step.  Explicitly, the state at the $n$-th step  is given by  
            \begin{align}\label{staten-1}  
                \widetilde {\map C}_\theta^{n-1}    \Big(  |j,j\>\<j,j|_{\st n}\Big)    =   \sum_{m=   j-n+1}^j \,    p(n-1,m,\theta)  \,   |j,m\>\<j,m|_{\st n}  \, , 
            \end{align}  
        where $p(n-1,m,\theta)$ is the probability distribution given by 
           \begin{align}\label{recycling-distribution}
        	p(n,m,\theta)=&\nonumber \sum_{i=j-m}^{n}(-1)^{i+j-m}
        	\begin{pmatrix}
        		n \\ i
        	\end{pmatrix}
            \begin{pmatrix}
        		i \\ j-m
        	\end{pmatrix}\frac{i!}{(2j)^{i}}\\
        	\nonumber=&(-1)^{j-m+1}\,  \frac{2j}{(1-\cos\theta)}  \,  \frac{n!}{(n-j+m)!}      \\
        	&     \times U\left(j-m+1,n+2,-\frac{2j}{1-\cos\theta} \right) \, ,
        \end{align}
       $U$ being Kummer's function.    
        
        To get the longevity we need to calculate $F(j,\theta,n)$ by
        \begin{align}\label{(equ-D1)}
        	F(j,\theta,n) =\sum_m  \,  p(n-1,m,\theta)    ~  F(j,\theta,m) \ ,
        \end{align}
        where $F(j,\theta,m)$ is the fidelity when using $|j,m\>_{\st n}$ as program state.
        
        The average fidelity $F(j,\theta,m)$ can be computed in terms of  the entanglement fidelity, using the relation 
        \begin{align}
        	F(j,\theta,m)=\dfrac{1}{3}+\dfrac{2}{3}F^{\rm (e)}(j,\theta,m) \ ,
        \end{align}
       Using  Eq. (\ref{optimalU}), the entanglement fidelity can be evaluated explicitly as 
             \begin{align}
            	F^{\rm (e)}(j,\theta,m)= 1- \dfrac{(1+2j-2m)(1-\cos\theta)}{2j}+O\left(\frac{1}{j^{2}}\right) \ . 
            \end{align}
        Going back to the average fidelity, we obtain 
            \begin{align}\label{equ-D4}
            	F(j,\theta,m)=1- \dfrac{(1+2j-2m)(1-\cos\theta)}{3j}+O\left(\frac{1}{j^{2}}\right) \ .
            \end{align}

      \begin{figure}[ht]
        	\centering
        	\includegraphics[width=0.51\textwidth]{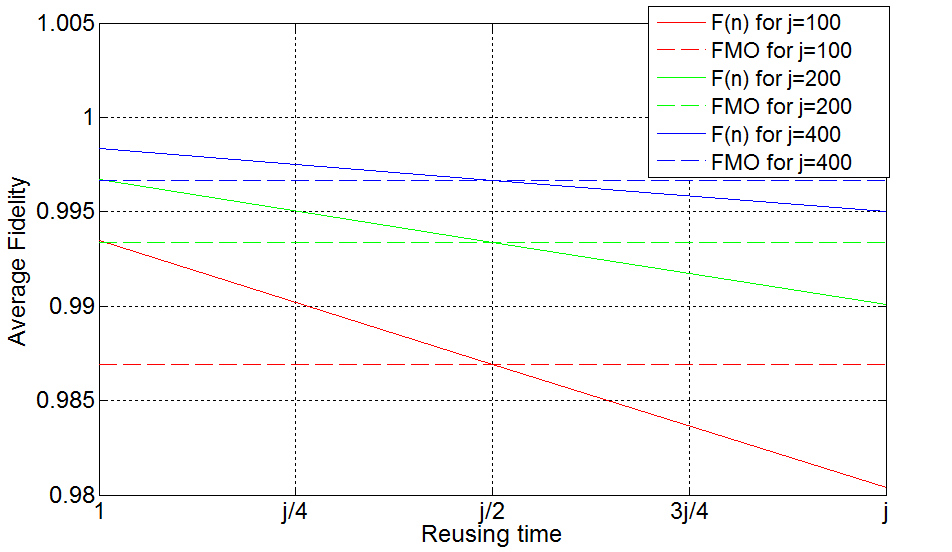}
        	\caption{\footnotesize
        		\textbf{Degradation of the fidelity with the number of recycling steps.}   The dependence of the fidelity on the number  $n$ of recycling steps is plotted for $j=100$  (red solid line),   $j=200$  (green solid line), and $j =400$ (blue solid line), in the case of rotation angle $\theta=  \pi$.    The plot shows an inverse linear scaling  with the recycling step $n$.     The dotted lines give the values of the MO fidelities for $j=100$  (red), $j=200$  (green), and $j=400$ (blue).   The  fidelity of this protocol falls under the MO fidelity when the number of recycling steps is larger than $j/2$. }
        	\label{fig5}
        \end{figure}
  The exact dependence of  the fidelity on $n$ is shown  in Figure \ref{fig5} for different values of the spin and for rotation angle $\theta= \pi$. 
        Interestingly, the longevity is exactly equal to the asymptotic value $j/2$ for all the values of $j$ shown  in the figure.

        In asymptotics of $j\rightarrow\infty$, by using recursion formula of Kummer's $U$ function\\
        \begin{align}
            \nonumber U(a,b,z)=&(2a-b+z+2)U(a+1,b,z)
        	\\&-(a+1)(a-b+2)U(a+2,b,z) \, ,
        \end{align}
        we will finally get
        \begin{align}\label{equ-D8}
        	p(n,m,\theta)\nonumber=& \dfrac{2j}{n(1-\cos\theta)+2j}\cdot   \left[\dfrac{n(1-\cos\theta)}{n(1-\cos\theta)+2j}\right]^{j-m}\\
        	& + O\left(\frac{1}{j}\right) \, .
        \end{align}
         One can see directly that in asymptotics, $F(j,\theta,m)$ is a arithmetic progression and $p(n,m,\theta)$ is a geometric progression.
       Inserting the above  expressions into Eq. (\ref{(equ-D1)})   we obtain  
            \begin{align}\label{recyclingfidelity}
            F(j,\theta,n) = 1 - \dfrac{1-\cos\theta}{3j}\cdot\dfrac{n(1-\cos\theta)+j}{j} + O\left(\frac{1}{j^{2}}\right)   \, .
            \end{align}
        Comparing with the MO fidelity, we obtain that the longevity of the quantum advantage  tends to $L(j,\theta)=j/(1-\cos\theta)$.
             
  We showed the explicit  calculation of  $F(j,\theta,m)$ and $p(n-1,m,\theta)$ when the interaction time is fixed at every step. 
    More general strategies where  the interaction time is  optimized at every step can be studied in the same way.  In the large $j$ limit, we find that such step-by-step optimization is not needed:  the  fidelity  tends to the same value, no matter whether the interaction time is optimized at every step or once for all. As a result, the longevity is the same in both scenarios.

    \section{Controlling  spin-$k$ particles}\label{app:spink}
        Following the structure of the optimal control mechanism for  spin $1/2$,  we choose the program state to be $|j,j\>_{\bf n}$ and we let the two spins undergo  the unitary gate
         \begin{align}
        U_\theta^{(k)}  =     \exp  \left[    -i \theta \, \frac{  \,   2\st J \cdot \st K}{2j+1}  \right] \, .
        \end{align} 
Using the above strategy,  we can explicitly compute the  entanglement fidelity, given by 
        \begin{align}\label{D2}
        &F^{\rm (e)}(j,k,\theta)\nonumber= \int d{\bf n}  ~F_{\rm{e}}(j,k,\theta, \bf{n})\\
        =\int d{\bf n}&\<\Phi^{(k)+}_{\theta,{\bf{n}}}|\Tr_{\rm control}\left[(\map U_
        {\theta}\otimes \map I)(\phi_{\bf{n}}^{(j)} \otimes \Phi^{(k)+})\right]|\Phi^{(k)+}_{\theta,{\bf{n}}}\> \, .
        \end{align}
        Here $\Phi^{(k)+}$ denotes the projector on the canonical maximally entangled state   and  $|\Phi^{(k)+}_{\theta,{\bf{n}}}\>$    is the rotated maximally entangled state defined by 
        \begin{align}|\Phi^{(k)+}_{\theta,{\bf{n}}}\>:   = (  V_{\theta, \bf n}^{(k)}  \otimes I)   |\Phi^{(k)+}\>  \, .
        \end{align}
        
Putting the formula of $U_{\theta}^{(k)}$ in Eq. (\ref{D2}), using the expressions of the Clebsch-Gordan coefficients, 
        we  arrive to the asymptotic expression 
        \begin{align}
        F^{\rm (e)}(j,k,\theta)= 1-\dfrac{2k(k+1)}{3j}(1-\cos\theta) \ + \ O\left(\frac{1}{j^{2}}\right) \, .
        \end{align}
  The average fidelity is then given by
        \begin{align}
        F(j,k,\theta)= 1-\dfrac{k(2k+1)}{3j}(1-\cos\theta) \ + \ O\left(\frac{1}{j^{2}}\right) \, .
        \end{align}
        
      A similar calculation can be done for the  MO strategy consisting in measuring the direction with  the coherent state POVM and then performing the conditional operation  $V_{\theta, {\bf n'}}^{(k)}$ on the target.  
      The entanglement fidelity of this strategy is given by     \begin{align}
        F^{\rm (e)}_{\rm MO}(j,k,\theta)\nonumber
        =\int d{\bf n}\int d{\bf n'}\Tr\left[P_{\bf n'}\phi_{\bf n}\right]F^{\rm (e)}(j,k, \theta, {\bf n}, {\bf n'}) 
        \end{align}
        with \begin{align}
        F^{\rm (e)}(j,k,\theta, {\bf n}, {\bf n'}) =
        \dfrac{1}{(2k+1)^{2}}\left|\Tr\left[V_{\theta,{\bf n}}^{\dagger}V_{\theta,{\bf n'}}^{(k)}\right]\right|^{2} \, .
        \end{align}
        
        By denoting $\varphi$ as the angle between axis $\bf n$ and $\bf n'$, and by  $\tau$ the rotation angle for the rotation   $V_{\theta,{\bf n}}^{\dagger}V_{\theta,{\bf n'}}^{(k)}$, the entanglement fidelity can be rewritten as 
         \begin{align}
        F^{\rm (e)}_{\rm MO}(j,k,\theta)=\dfrac{\int_{0}^{\pi}d\varphi \sin\varphi (\cos\varphi)^{4j}\dfrac{\sin^{2}(\frac{2k+1}{2}\tau)}{\sin^{2}\frac{\tau}{2}}}{(2k+1)^{2}\int_{0}^{\pi}d\varphi \sin\varphi (\cos\varphi)^{4j}} \, . 
        \end{align}
Performing the average, we obtain the asymptotic expression        \begin{align}
        F^{\rm (e)}_{\rm MO}(j,k,\theta) = 1-\dfrac{4k(k+1)}{3j}(1-\cos\theta) +O\left(\frac{1}{j^{2}}\right) \, ,
        \end{align}
        which can then be used to evaluate the average fidelity as
        \begin{align}
        F_{\rm MO}(j,k,\theta) = 1-\dfrac{2k(2k+1)}{3j}(1-\cos\theta) +O\left(\frac{1}{j^{2}}\right) \ .
        \end{align}

\end{document}